\documentclass[a4paper, superscriptaddress, amsfonts, amssymb, amsmath, reprint, showkeys, nofootinbib, twoside]{revtex4-1}
\usepackage[english]{babel}
\usepackage[utf8]{inputenc}
\usepackage{xcolor}
\usepackage{bm}
\usepackage{graphicx}
\usepackage{float}
\usepackage{amsmath}
\usepackage{amsfonts,amssymb,amsthm}
\usepackage[colorinlistoftodos, color=green!40, prependcaption]{todonotes}
\usepackage{amsthm}
\usepackage{mathtools}
\usepackage{physics}
\usepackage{xcolor}
\usepackage{graphicx}
\usepackage[left=23mm,right=13mm,top=35mm,columnsep=15pt]{geometry} 
\usepackage{adjustbox}
\usepackage{placeins}
\usepackage[T1]{fontenc}
\usepackage{lipsum}
\usepackage{csquotes}
\usepackage[pdftex, pdftitle={Article}, pdfauthor={Author}]{hyperref} 

\begin{document}
\title{Partial information decomposition for mixed discrete and continuous \\ random variables}

\author{Chiara Barà}
    \affiliation{Department of Engineering, University of Palermo, Palermo, Italy}
\author{Yuri Antonacci}
    \affiliation{Department of Engineering, University of Palermo, Palermo, Italy}
\author{Marta Iovino}
    \affiliation{Department of Engineering, University of Palermo, Palermo, Italy}
\author{Ivan Lazic}
    \affiliation{Faculty of Technical Sciences, University of Novi Sad, Novi Sad, Serbia}   
\author{Luca Faes}
    \email[Correspondence email address: ]{luca.faes@unipa.it}
    \affiliation{Department of Engineering, University of Palermo, Palermo, Italy}

\begin{abstract}
The framework of Partial Information Decomposition (PID) unveils complex nonlinear interactions in network systems by dissecting the mutual information (MI) between a target variable and several source variables. While PID measures have been formulated mostly for discrete variables, with only recent extensions to continuous systems, the case of mixed variables where the target is discrete and the sources are continuous is not yet covered properly. Here, we introduce a PID scheme whereby the MI between a specific state of the discrete target and (subsets of) the continuous sources is expressed as a Kullback-Leibler divergence and is estimated through a data-efficient nearest-neighbor strategy. The effectiveness of this PID is demonstrated in simulated systems of mixed variables and showcased in a physiological application. Our approach is relevant to many scientific problems, including sensory coding in neuroscience and feature selection in machine learning.
\end{abstract}


\maketitle

The framework of Partial Information Decomposition (PID) was originally introduced in the seminal work by Williams and Beer~\cite{williams2010nonnegative} with the aim of decomposing the mutual information (MI) between a target variable and a set of several source variables into separate unique, redundant and synergistic contributions, so as to reveal modes of interaction that would remain hidden when considering the individual MIs between the target and each source. Since its inception, the PID framework has gained big popularity as a tool for investigating high-order interdependencies among correlated variables in several fields of science, including neuroscience, physiology and biology, sociology, climatology, and machine learning \cite{wibral2017partial, sherrill2021partial,faes2021information,varley2022untangling,wollstadt2023rigorous,stramaglia2024disentangling}.
At the same time, the framework has undergone a continuous development related to the fact that the PID problem cannot be solved using the classical Shannon theory of Information, so that novel definitions of one of the PID terms (typically, the redundant term) have been formulated~\cite{harder2013bivariate,bertschinger2014quantifying,barrett2015exploration,ince2017measuring}.

A related issue with great practical importance, which is the focus of the present letter, is that both the original formulation of the PID framework~\cite{williams2010nonnegative} and many of its subsequent developments \cite{harder2013bivariate,ince2017measuring,bertschinger2014quantifying} are valid for variables with discrete alphabets. This restriction limits the applicability of the majority of PID measures to categorical data or quantized representations of continuous-valued data. 
Alternatively, some PID formulations working for continuous random variables have been later advanced, though they are limited by specific assumptions on the data distribution (e.g., the Gaussian distribution~\cite{barrett2015exploration}) or based on specific operational definitions (e.g., decision-theoretic~\cite{pakman2021estimating}  or game-theoretic~\cite{ince2017measuring}); only recently, a continuous PID measure drawing only on information-theoretic concepts has been designed together with an efficient estimation scheme~\cite{ehrlich2024partial}. 
Yet, formulations of the PID for mixed random variables where the target variable is discrete and the sources are continuous-valued are not popular, although they would be highly desired in several applications, e.g., in machine learning or computational physiology; for instance, contexts such as feature selection, where a discrete representation of the class variable is needed while maintaining a continuous representation of the features \cite{wollstadt2023rigorous}, or neuroscience paradigms, where discrete stimuli are conveyed by nervous systems generating continuous-valued responses \cite{borst1999information}, would benefit from the availability of a PID applicable to mixed discrete and continuous variables.

Here, we aim to address this issue by proposing a non-parametric procedure, based on nearest-neighbor entropy estimation~\cite{kraskov2004estimating}, for the decomposition of the MI between a discrete target variable $Y$ and $n$ continuous source variables $\textbf{X}=\{X_1,\ldots, X_n\}$, respectively taking values in the discrete alphabet $\mathcal{A}_Y=\{1,\dots,Q\}$ and in the continuous domain $\mathcal{D}_{\textbf{X}} \subseteq \mathbb{R}^n$. 
We begin observing that the PID can be formulated in a compact form by expanding the MI between target and sources as:
\begin{equation}
    I(Y;\textbf{X}) = \sum_{j=1}^n U(Y;X_j)+R(Y;\textbf{X})+S(Y;\textbf{X}),
    \label{coarsegrainedPID}
\end{equation}
where $U(Y;X_j)$, $R(Y;\textbf{X})$, and $S(Y;\textbf{X})$ denote respectively the \textit{unique information} shared by \textit{Y} exclusively with the $j^{\mathrm{th}}$ source, the \textit{redundant information} concomitantly shared by all sources with \textit{Y}, and the \textit{synergistic information} emerging only when all the sources are jointly observed.
It is important to note that the original formulation of the PID is more articulate, as it comprises exclusively the terms appearing in (\ref{coarsegrainedPID}) only when $n=2$ source variables are considered. In fact, exploiting a mathematical lattice structure to represent the set-theoretic intersection of multiple variables, the information shared by the target with all sources can be decomposed into partial information (PI) atoms defined over the nodes of the lattice:
\begin{equation}
    I(Y;\textbf{X}) = \sum_{\alpha \in \mathcal{S}(\textbf{X})} I_{\delta}(Y;\alpha),
    \label{PID}
\end{equation}
where $\mathcal{S}(\textbf{X})$ is the collection of all subsets of sources such that no source is a superset of any other~\cite{williams2010nonnegative}. The atoms of the lattice over which the PI is defined can be identified and ordered for arbitrarily many source variables according to a specific notion of parthood~\cite{gutknecht2021bits}; for $n=2$ sources the PI atoms correspond to the PID terms in (\ref{coarsegrainedPID}): $I_{\delta}(Y;\{X_j\})=U(Y;X_j)$, $I_{\delta}(Y;\{X_1\}\{X_2\})=R(Y;\textbf{X})$, and $I_{\delta}(Y;\{X_1,X_2\})=S(Y;\textbf{X})$. However, since the number of PID atoms $m=|\mathcal{S}(\textbf{X})|$ grows super-exponentially with $n$ (e.g., $m=18$ when $n=3$ and $m=166$ when $n=4$), there is the need to sum the PI of some of the $m$ atoms in (\ref{PID}) to retrieve the $n+2$ quantities in (\ref{coarsegrainedPID}); here we do this according to the coarse-graining procedure proposed by Rosas et al.~\cite{rosas2020reconciling} (see suppl. material, Sect. I). 

As early noted in ~\cite{williams2010nonnegative}, the PID problem is underdetermined because the number of PI atoms to be computed always exceeds the number of available classical MI quantities providing constraints through (\ref{PID}) (i.e., $m>2^n-1$). Therefore, besides the lattice structure of the information atoms and their ordering, to complete the PID it is necessary to define a so-called \textit{redundancy function} $I_\cap(Y;\alpha)$ over the lattice. This redundancy generalizes the MI and fulfills $I_\cap(Y;\alpha)=\sum_{\beta \preceq \alpha}I_\delta(Y;\beta)$, so that knowing the redundancy function allows to retrieve the PI of each atom recursively or through a M\"{o}bius inversion:
\begin{equation}
    I_\delta(Y;\alpha) = I_\cap(Y;\alpha) - \sum_{\beta \prec \alpha} I_\delta(Y;\beta),
    \label{partial_information_definition}
\end{equation}
where the relations identified by $\prec$ are determined according to the partial ordering imposed by the lattice structure (see suppl. material, Sect. I).

In this work we employ the redundancy function defined by William and Beer~\cite{williams2010nonnegative}, which exploits a measure of the \textit{specific information} shared by the sources with a particular observation of the discrete target variable:
\begin{equation}
    I_\cap(Y;\alpha) = \sum_{i\in A_Y} p(Y=i) \min_{\{X_s\}\in\alpha}I\left(Y=i;X_s \right),
    \label{redundacy_definition}
\end{equation}
where the information brought redundantly to the target by the sources $\{X_s\}$ collected in the atom $\alpha$ of the lattice is defined taking the minimal information provided by any of these sources to each specific outcome $i$ of the target. The specific MI naturally lends itself to be generalized to the case of discrete target and continuous sources, for which it can be expressed as:
\begin{equation}
    I(Y=i;X_s) = \int_{x_s\in D_{X_s}}p(x_s|i) \ln\frac{p(x_s|i)}{p(x_s)}dx_s,
    \label{spec_inf}
\end{equation}
where $p(x_s)$ and $p(x_s|i)$ denote the probability density of $X_s$ when considering, for the variable $Y$, all possible outcomes or a specific outcome $i$, respectively; note that in (\ref{spec_inf}) we use the natural logarithm, as typically done for continuous variables.

\begin{figure}
    \centering
    \includegraphics[scale = 0.9]{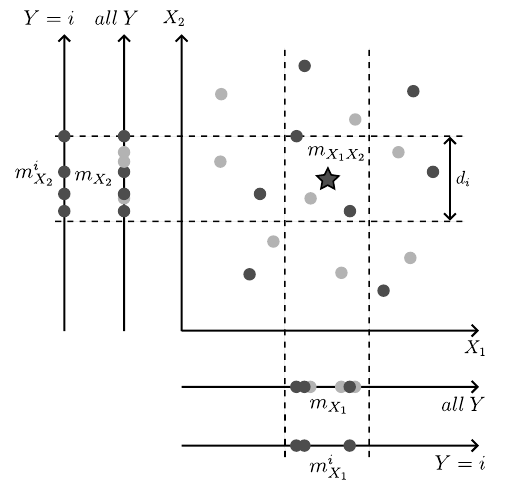} 
    \caption{Schematic representation of the nearest-neighbor estimation of the specific MI (\ref{spec_inf}) for a binary target variable $Y$ and two continuous sources $X_1$ and $X_2$. The estimation proceeds by computing the distance (here, the Chebyshev distance) between each reference sample $\mathbf{x}^i$ (star symbol) and its $k^\mathrm{th}$ neighbor (here, $k=2$) counted among the samples with $Y=i$ (black circles), and then counting the number of samples falling within this distance in the two-dimensional space ($m_{X_1X_2}=3$) and in the projected one-dimensional spaces considering all samples ($m_{X_1}=6, m_{X_2}=7$) or only the samples for which $\{Y=i\}$ ($m^i_{X_1}=3, m^i_{X_2}=4$).} 
    \label{fig:stima}
\end{figure}

The specific MI is estimated here expressing Eq. (\ref{spec_inf}) as the Kullback-Leibler divergence between the probabilities $p(x_s|i)$ and $p(x_s)$ and computing the relevant entropy and cross-entropy terms via the Kraskov-Stögbauer-Grassberger (KSG) nearest-neighbor MI estimator~\cite{kraskov2004estimating}.
The KSG strategy is here implemented performing a neighbor search among the samples of all sources associated with $\{Y=i\}$ to compute the entropy of $p(\textbf{x}|i)$, and a range search among all samples to compute the cross-entropy between $p(\textbf{x}|i)$ and $p(\textbf{x})$. This results in computing the specific MI for all sources as:
\begin{equation}
    \hat{I}(Y=i;\textbf{X}) = \psi(N)- \psi(N_i) + \psi(k) - \frac{1}{N_i} \sum_{\{\textbf{x}^i\}}\psi(m_{\textbf{X}}+1),
    \label{spec_inf_est_allsources}
\end{equation}
where the sum is extended to all the observations of $\textbf{X}$ associated with the specific outcome $\{Y=i\}$, denoted as $\{\textbf{x}^{i}\}$; in (\ref{spec_inf_est_allsources}), $\psi \left(\cdot \right)$ is the digamma function, $N$ is the total number of samples, $N_i$ the number of samples associated to the outcome $\{Y=i\}$, and $m_{\textbf{X}}$ is the number of observations of $\textbf{X}$ found at distance smaller than $d_i/2$ from the  observation $\textbf{x}^i$, with $d_i$ being the distance between $\textbf{x}^i$ and its $k^{\mathrm{th}}$ nearest neighbor.
Applying again the KSG strategy, the specific MI estimated for subsets of source variables, $X_s\subset \textbf{X}$, is obtained performing range searches within the range of distances $d_i$ determined in the highest dimension: 
\begin{equation} 
    \begin{split}
        & \hat{I}(Y=i;X_s) = \psi(N)- \psi(N_i) \\
        & + \frac{1}{N_i} \sum_{\{x^i_s\}}\biggl( \psi(m_{X_s}^i+1) - \psi(m_{X_s}+1)\biggr),
    \end{split}
    \label{spec_inf_est}
\end{equation}
where $m_{X_s}$ and $m_{X_s}^i$ are the number of observations of $X_s$ counted at distances smaller than $d_i/2$ from $x_s$, respectively considering all the samples or only the samples associated with $\{Y=i\}$; the sum is again extended to all observations of $X_s$ associated with $\{Y=i\}$.
The estimation procedure is illustrated in Fig. \ref{fig:stima}, and the complete derivations for $n=2$ and $n=3$ sources are provided in the suppl. material (Sect. II).

\begin{figure}
    \centering
    \includegraphics[scale = 0.97]{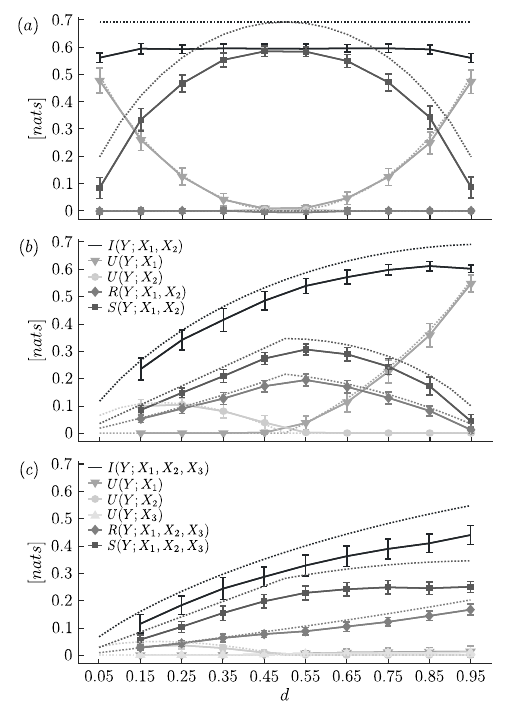} 
    \caption{Theoretical values (dotted lines) and estimated distributions (solid lines, mean $\pm$ std. dev. over 100 simulations) of the PID measures obtained for the XOR gate (\textit{a}), and for the AND gates with 2 (\textit{b}) or 3 sources (\textit{c}).}
    \label{fig:sim}
\end{figure}

The proposed estimator is validated in a simulation of two independent continuous variables with uniform distribution, $X_1 \sim \mathcal{U}(\textit{a},\textit{b})$ and $X_2 \sim \mathcal{U}(\textit{c},\textit{d})$, whose signs are combined through the Heaviside function ($\Theta(x)=1$ if $x\geq 0$, $\Theta(x)=0$ if $x<0$) to obtain the output of the Boolean XOR and AND gates, respectively as $Y=\Theta\left(X_1\cdot X_2 \right)$ and $Y=\Theta\left(X_1\right) \cdot \Theta\left(X_2\right)$. The case with three sources is also simulated considering a third input variable $X_3 \sim \mathcal{U}(\textit{a},\textit{b})$ in the AND configuration, i.e., $Y=\Theta\left(X_1\right) \cdot \Theta\left(X_2\right) \cdot \Theta\left(X_3\right)$.
These systems are analyzed by fixing $b=-a=0.5$ and varying $d$ with $c=d-1$: the theoretical trends reported in Fig. \ref{fig:sim} document that the MI is constant for the XOR gate and increases with $d$ for the AND gates~\cite{timme2014synergy}. The XOR gate yields exclusive synergy when $X_1$ and $X_2$ are positive or negative with the same probability ($d=0.5$; in this case both sources must be known to determine the output of the gate), while it yields exclusive unique contribution from $X_1$ when $X_2$ takes only negative ($d=0$) or only positive ($d=1$) values thus becoming irrelevant. 
In the AND gates the target is deterministic if $d=0$, but increasing $d$ the sources share more information with $Y$; such information is both redundant and synergistic, with synergy prevailing over redundancy, and is completed by unique contributions from $X_2$ for $d<0.5$ and from $X_1$ (and also $X_3$ for the case of three sources) for $d>0.5$.
Fig. \ref{fig:sim} reports also the trends of the PID terms estimated using $k=5$ neighbors over 300-point realizations of the variables, showing how the estimates follow reliably the true values of each measure, though exhibiting a bias that is more evident for the terms involving samples counted in higher dimensional spaces (i.e., the overall MI and the synergy). A detailed analysis (suppl. material, Sect. III) shows that the estimation bias is larger for higher-dimensional systems, and increases with a reduction of the length $N$ of the data and an increase of the number $k$ of neighbors searched, while the standard deviation increases at decreasing $N$ and is rather stable at varying $k$.

\begin{figure}
    \centering
    \includegraphics[scale = 0.97]{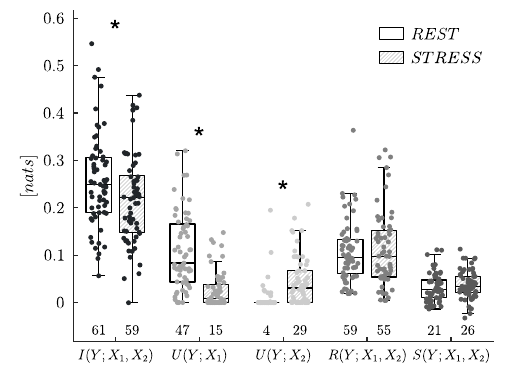} 
    \caption{Boxplot distributions and individual values of the PID applied to the MI between the binary target variable $Y$ measuring inspiration/expiration and the continuous source variables $X_1$  and $X_2$ measuring respectively heart period and systolic arterial pressure, computed in conditions of resting state and postural stress. *, $p<0.05$ paired Student’s \textit{t}-test, REST vs. STRESS. The number of subjects for which each measure is statistically significant according to surrogate data analysis is shown below each boxplot.}
    \label{fig:appl}
\end{figure}

As an applicative example, we use the physiological database in~\cite{javorka2018towards} to investigate the interplay between cardiovascular dynamics and respiratory activity in two widely studied physiological conditions, i.e., rest and postural stress~\cite{eckberg2009point}. 
The analyzed variables are the respiratory phase, taken as binary target variable $Y$ with symbols corresponding to \textit{inspiration} and \textit{expiration}, and the heart period and systolic arterial pressure, taken as continuous sources and measured as the sequences of pairs of consecutive cardiac interbeat intervals ($X_1$) and systolic pressure values ($X_2$). These variables were measured ($N=300$ observations) from the respiratory belt, electrocardiogram and finger arterial pressure signals in 61 subjects analyzed in the two conditions~\cite{javorka2018towards}.
The computation of the MI and the PID terms documents significant interactions between the cardiovascular dynamics and the respiratory phase (Fig. \ref{fig:appl}). The statistical significance of each PID term is assessed through a surrogate data approach based on random shuffling of the target samples, computation of the specific MI for each atom of the lattice, and statistical test based on percentiles with correction for multiple comparisons (Suppl. material, sect. IV).
The comparison of the two conditions indicates a weakening of the interaction between the cardiovascular and respiratory systems with postural stress, documented by the statistically significant decrease of $I(Y;X_1,X_2)$. Application of the PID reveals physiologically meaningful modulations induced by the change of posture, such as the dampening of the mechanism of respiratory sinus arrhythmia~\cite{eckberg2009point} reflected by the drop of $U(Y;X_1)$, and the enhancement of the cyclic intrathoracic pressure changes and the baroreflex mechanism related to ventilation~\cite{draghici2016physiological} reflected by the rise of $U(Y;X_2)$. Even if no variations are obtained for the redundant and synergistic terms, the prevailing redundant contribution in both conditions is likely related to the influence of ventilatory activity on both cardiac and vascular dynamics with similar latency~\cite{draghici2016physiological,eckberg2009point}.

The above results demonstrate the validity of our approach for the computation of the unique, redundant and synergistic contributions of continuous source variables to the information brought by a discrete target variable, showing how it reflects expected variations of the PID terms in theoretical systems as well as its feasibility over short-length realizations of multivariate processes.
Given the growing popularity of the PID framework in a variety of real-world contexts, we expect that the proposed estimator will find utilization in several real-world scientific problems usefully modeled by mixed discrete and continuous random variables.

\renewcommand{\theequation}{S.\arabic{equation}}
\renewcommand{\thetable}{S.\arabic{table}}
\renewcommand{\thefigure}{S.\arabic{figure}}
\setcounter{equation}{0}
\setcounter{table}{0} 
\setcounter{figure}{0} 

\section*{Supplementary material}

\section{PID in multivariate systems \label{coasegraining}}
\begin{figure*}
    \centering
    \includegraphics[scale=0.95]{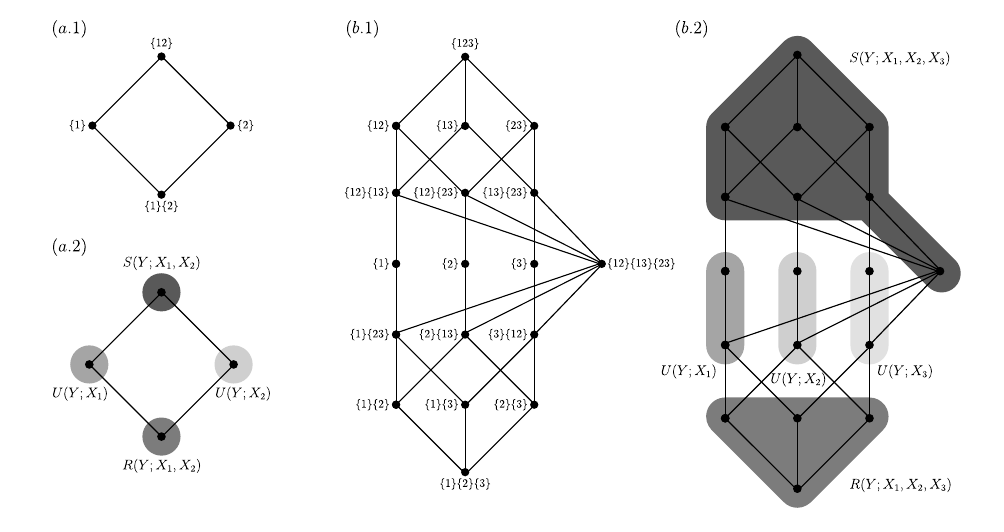}
    \caption{Redundancy lattices representing the PI atoms ordering (panels \textit{.1}) and the PID components definition (panels \textit{.2}) for the PID with two sources (panels \textit{a.}) and for the PID with three sources (panels \textit{b.}).}
    \label{fig:cg}
\end{figure*}

The redundancy lattice orders specific sources (or collections of them) according to the information they provide to a target variable; in particular, atoms placed higher in the structure provide at least as much redundant information as those below ~\cite{williams2010nonnegative}.
In the simplest case of three variables (i.e., one target and two sources), the lattice consists of the four partial information (PI) atoms depicted in Figure~\ref{fig:cg}(\textit{a.1}). By using the common notation for which redundant information is provided by sources in different brackets and synergistic information by sources in the same brackets, each atom of this structure gives a specific contribution to the mutual information $I(Y;\textbf{X})$, i.e., $R(Y;\textbf{X}) = I_\delta(Y;\{X_1\}\{X_2\})$, $U(Y;X_1) = I_\delta(Y;\{X_1\})$, $U(Y;X_2) = I_\delta(Y;\{X_2\})$, and $S(Y;\textbf{X}) = I_\delta(Y;\{X_1,X_2\})$ (see Figure~\ref{fig:cg}(\textit{a.2})).

The structural complexity of the redundancy lattice increases with the dimensionality of the system~\cite{gutknecht2021bits}. In the 4-variable lattice employed when working with one target and three sources, the eighteen PI atoms are linked as in Figure~\ref{fig:cg}(\textit{b.1}). The redundant information provided by the three sources is represented by the lowest atom (i.e., $\{1\}\{2\}\{3\}$), while the information redundantly associated to combinations of pairs of single sources (i.e., $\{1\}\{2\}$, $\{1\}\{3\}$ and $\{2\}\{3\}$), of single sources and couples of them (i.e., $\{1\}\{23\}$, $\{2\}\{13\}$ and $\{3\}\{12\}$), and of couples of sources (i.e., $\{12\}\{13\}$, $\{12\}\{23\}$, $\{13\}\{23\}$ and $\{12\}\{13\}\{23\}$) are associated to nodes placed in higher positions in the lattice. Moreover, PI atoms representing the information provided individually by the sources (i.e., $\{1\}$, $\{2\}$ and $\{3\}$) are also present, while those attributed to synergistic combinations of these (i.e., $\{12\}$, $\{13\}$, $\{23\}$ and $\{123\}$) are placed in the highest positions of the redundancy lattice.

As described in detail in the Appendix of~\cite{rosas2020reconciling}, a coarse-graining approach can be introduced which merges several atoms and sums the corresponding partial information measures $I_\delta(Y;\alpha)$ to obtain single information values for each of the unique, synergistic and redundant components. Specifically, as graphically shown in Figure~\ref{fig:cg}(\textit{b.2}), by using the first-order coarse-grained  approach \cite{rosas2020reconciling}, the PID components are obtained summing the PI of the atoms collected in $\mathcal{R}=\{\{1\}\{2\}\{3\},\{1\}\{2\},\{1\}\{3\},\{2\}\{3\}\}$ for the redundancy term $R(Y;\textbf{X})$, in $\mathcal{S}=\{\{12\}\{13\}\{23\},\{12\}\{13\},\{12\}\{23\},\{13\}\{23\},\{12\},\\\{13\},\{23\},\{123\}\}$  for the synergistic term $S(Y;\textbf{X})$, and in $\mathcal{U}_1=\{\{1\},\{1\}\{23\}\}$, $\mathcal{U}_2=\{\{2\},\{2\}\{13\}\}$, $\mathcal{U}_3=\{\{3\},\{3\}\{12\}\}$ for the unique information $U(Y;X_1)$, $U(Y;X_2)$ and $U(Y;X_3)$, respectively.

\section{Estimation strategy}
Here we present in detail the estimation approach introduced in our work to decompose the information shared by continuous source variables and a discrete target variable via the PID framework. As reported in the main paper, the computation of the unique, redundant and synergistic information measures can be reduced to the estimation of the information specifically provided by the sources to a specific outcome of the target. 
Considering a target $Y$ with possible outcomes $i \in A_Y=\{1,\dots,Q\}$ and a variable $X_s$ representing a generic source or combination of $n_s$ sources from $\textbf{X}=\{X_1,\ldots,X_n\}$ assuming values in $D_{X_s} \subseteq \mathbb{R}^{n_s}$, the specific information shared by $X_s$ with $Y$ when considering only the outcome $i$, defined in Eq. (5) of the main paper, can be expressed as the Kullback-Leibler divergence between $p(x_s|i)$ and $p(x_s)$, indicative respectively of the probability densities of $X_s$ when considering, for the variable $Y$, all possible outcomes or a specific outcome $i$:
\begin{align} \label{Ispec_dec}
    I&(Y=i;X_s) = H_{X_s|i}(X_s) - H_{X_s|i}(X_s|i)\\\nonumber
    = &\mathbb{E}_{p_{x_s|i}}\left[\ln \frac{1}{p(x_s)}\right]-\mathbb{E}_{p_{x_s|i}}\left[\ln \frac{1}{p(x_s|i)}\right]= \nonumber \\\nonumber
    =&- \int_{x_s\in D_{X_s}}p(x_s|i) \ln p(x_s) dx_s \\\nonumber
    &+ \int_{x_s\in D_{X_s}}p(x_s|i) \ln p(x_s|i) dx_s.  
\end{align}
In (\ref{Ispec_dec}), the first term $H_{X_s|i}(X_s)$ represents the cross-entropy of the  unconditioned density $p(x_s)$ with respect to the conditional density $p(x_s|i)$, while the second term $H_{X_s|i}(X_s|i)$ represents the entropy of $p(x_s|i)$.

The implemented approach makes use of the nearest-neighbor estimation method and, as similarly done in \cite{ross2014mutual}, considers how the observations of the source variables are associated with specific outcomes of the target as a criterion for defining the searching space. This approach allows to disentangle the information shared between variables of different types, including both discrete and continuous variables. 
Specifically, here we employ the estimation strategy proposed by Kraskov-Stögbauer-Grassberger (KSG) ~\cite{kraskov2004estimating} to minimize the estimation bias resulting from the combination of entropy terms derived using a different number of observations and exploited to derive \textit{specific} and \textit{global} probabilities, i.e., $p(x_s|i)$ and $p(x_s)$. Indeed, even if the dimensionality of the searching space remains unchanged (e.g., while computing $H_{X_1|i}(X_1)$ and $H_{X_1|i}(X_1|i)$), the combination of cross-entropy and entropy estimates has comparable effects to those observed in the estimation of entropy terms which account for searching neighbors on spaces of different dimensions. 
Moreover, the proposed estimator considers the fact that solving the PID implies computing specific MIs for a different number of source variables (e.g., for the case of two sources, $I(Y=i;X_1,X_2)$, $I(Y=i;X_1)$ and $I(Y=i;X_2)$), and thus there is the need to apply the KSG strategy again, this time in a more standard way which accounts for the different dimension of the entropy terms to be computed. In fact, working with $n$ source variables, the application of the PID framework consists in the computation of the MI terms of the $n$ sources and of any possible combination (of dimension from 2 to $n$) with each of the $Q$ outcomes of the target.

The estimation strategy discussed above and graphically illustrated in Fig. (1) of the main paper can be reduced to the computation of four types of entropy measures. Specifically, with $d_i$ being the distance between an observation of $\textbf{X}$ associated with the specific outcome $\{Y = i\}$ and its $k^{th}$ nearest neighbor, estimation of the following four terms is needed: 
\begin{itemize}

\item the entropy of all the sources \textbf{X} conditioned to the specific state $\{Y = i\}$ of the target, computed for the conditional probability $p(\textbf{x}|i)$:
\begin{equation}
    \hat{H}_{\textbf{X}|i}(\textbf{X}|i) = -\psi(k)+\psi(N_i)+\frac{n}{N_i}\sum_{\{\textbf{x}^i\}} \ln d_i,
    \label{Hneighbor}
\end{equation}
where $\psi(\cdot)$ is the digamma function, $N$ the total number of samples and $N_i$ the number of samples associated with the outcome $\{Y = i\}$;

\item the cross-entropy of all the sources \textbf{X} estimated with respect to the specific state $\{Y = i\}$ of the target, computed between the unconditioned probability $p(\textbf{x})$ and conditional probability $p(\textbf{x}|i)$:
\begin{equation}
    \hat{H}_{\textbf{X}|i}(\textbf{X}) = \psi(N)+\frac{1}{N_i}\sum_{\{\textbf{x}^i\}} \big( n \ln d_i - \psi(m_{\textbf{X}}+1) \big),
    \label{Hrangehigh}
\end{equation}
where $m_\textbf{X}$ is the number of observations of $\textbf{X}$ found at distance smaller than $d_i/2$ from the observation $\textbf{x}^i$;

\item the entropy of the generic combination of sources $X_s$ conditioned to the specific state $\{Y = i\}$ of the target, computed for the conditional probability $p(x_s|i)$: 
\begin{equation}
    \hat{H}_{X_s|i}(X_s|i) = \psi(N_i)+\frac{1}{N_i}\sum_{\{x_s^i\}} \big( n_s \ln d_i - \psi(m_{X_s}^i+1) \big),
    \label{Hrangelowspecific}
\end{equation}
where $m_{X_s}^i$ is the number of observations of $X_s$ associated with $\{Y = i\}$ found at distance smaller than $d_i/2$ from the observation $x_s^i$;

\item the cross-entropy of the generic combination of sources $X_s$ estimated with respect to the specific state $\{Y = i\}$ of the target, computed between the unconditioned probability $p(x_s)$ and conditional probability $p(x_s|i)$:
\begin{equation}
    \hat{H}_{X_s|i}(X_s) = \psi(N)+\frac{1}{N_i}\sum_{\{x_s^i\}} \big( n_s \ln d_i - \psi(m_{X_s}+1) \big),
    \label{Hrangelowglobal}
\end{equation}
where $m_{X_s}$ is the number of observations of $X_s$ found at distance smaller than $d_i/2$ from the observation $x_s^i$.
\end{itemize}
Importantly, only the entropy computed for the smallest subset of points (i.e., the specific entropy of $\textbf{X}$ given $\{Y = i\}$ of Eq. (\ref{Hneighbor}), calculated in the highest-dimensional space) is estimated via a neighbor search, while the other entropy terms are computed via range searches performed in the same highest-dimensional space (Eq. (\ref{Hrangehigh}), considering all points) or in the lower-dimensional spaces populated by the realizations of $X_s$ (either specific for $\{Y = i\}$, Eq. (\ref{Hrangelowspecific}), or global for any value of $Y$,  Eq. (\ref{Hrangelowglobal})).

The specific information terms necessary for applying the PID decomposition can be computed combining the entropy terms listed above as indicated in Eq. (\ref{Ispec_dec}). 
In particular, the equations for the specific mutual information quantities assessed in the case of $n=2$ and $n=3$ are reported as follows.

\paragraph*{\textbf{Two source variables framework}}
\hspace{1 mm}  Considering two continuous source variables $X_1$ and $X_2$, the distances $d_i$ to the $k^{th}$ neighbor are evaluated for the realizations of $\textbf{X} = [X_1,X_2]$ associated with a specific output $i$ of the discrete variable $Y$, i.e., $\{(x_1,x_2)^i\}$. The specific information measure relevant to the whole source space is obtained as:
\begin{align}
    I&(Y=i;X_1,X_2) = \psi(N)-\psi(N_i)+\psi(k) \\ \nonumber
    &-\frac{1}{N_i}\sum_{\{(x_1,x_2)^i\}}  \psi(m_{X_1X_2}+1),
\end{align}
while the specific mutual information terms for each single source $X_j$, $j \in \{1,2\}$, are obtained as:
\begin{align}
    I&(Y=i;X_j) = \psi(N)-\psi(N_i)\\ \nonumber
    &+\frac{1}{N_i}\sum_{\{x_j^i\}}  \big( \psi(m_{X_j}^i+1)-\psi(m_{X_j}+1) \big).
\end{align}

\paragraph*{\textbf{Three source variables framework}}
\hspace{1 mm}  Following the same estimation approach, the measures relevant the information specifically shared by a specific outcome of the target variable $Y$ and the source variables for the evaluation of all the partial information terms characterizing a four-variables lattice can be obtained. 
Specifically, for each possible outcome of the target $i$, the specific mutual information measure is firstly obtained in the higher dimensional space spanned by the realizations of $\textbf{X} = [X_1,X_2,X_3]$ as:
\begin{align}
    I&(Y=i;X_1,X_2,X_3) = \psi(N)-\psi(N_i)+\psi(k)\\ \nonumber
    &-\frac{1}{N_i}\sum_{\{(x_1,x_2,x_3)^i\}}  \psi(m_{X_1X_2X_3}+1),
\end{align}
and then, by using the same searching distances, in the lower dimensional spaces accounting for any couple of sources $X_j$ and $X_w$, $\{j,w\} \in \{\{1,2\},\{1,3\},\{2,3\}\}$:
\begin{align}
    I&(Y=i;X_j,X_w) = \psi(N)-\psi(N_i)\\ \nonumber
    &+\frac{1}{N_i}\sum_{\{(x_j,x_w)^i\}}  \big( \psi(m_{X_jX_w}^i+1)-\psi(m_{X_jX_w}+1) \big),
\end{align}
or accounting for any individual source $X_j$, $j \in \{1,2,3\}$:
\begin{align}
    I&(Y=i;X_j) = \psi(N)-\psi(N_i)\\ \nonumber
    &+\frac{1}{N_i}\sum_{\{x_j^i\}}  \big( \psi(m_{X_j}^i+1)-\psi(m_{X_j}+1) \big).
\end{align}

\section{Evaluation of the estimation bias and standard deviation}
When working in simulated settings, the theoretical values of the MI decomposition quantities are known, allowing to evaluate the reliability (in particular, the bias) of the estimates obtained using the proposed approach. In this section, the estimation bias and the estimation variance are investigated at varying the number \textit{N} of observations of the involved variables and the estimation parameter \textit{k}, as well as increasing the dimensionality of the system, i.e., the number of source variables \textit{n}, for the simulations reported in the main paper. 

Considering three independent uniformly distributed continuous variables, $X_1 \sim \mathcal{U}(\textit{a},\textit{b})$, $X_2 \sim \mathcal{U}(\textit{c},\textit{d})$ and $X_3 \sim \mathcal{U}(\textit{a},\textit{b})$, with $a=-b=-0.5$, $d$ in the range $\left[0.05:0.1:0.95 \right]$ and $c=d-1$, the Boolean logic gate of XOR with two sources (i.e., XOR2) and the Boolean logic-gate AND considering both two (i.e., AND2) and three (i.e., AND3) source variables are simulated respectively as $Y=\Theta\left(X_1 \cdot X_2\right)$, $Y=\Theta\left(X_1\right) \cdot \Theta\left(X_2\right)$ and $Y=\Theta\left(X_1\right) \cdot \Theta\left(X_2\right) \cdot \Theta\left(X_3\right)$, with the Heaviside function $\Theta(x)=1$ if $x\geq 0$ and $\Theta(x)=0$ if $x<0$. 
The probabilities of the three sources of being positive or negative are theoretically known and, exploiting their statistical independence, also the joint probabilities can be derived as reported in tables \ref{prob_th_2sources} and \ref{prob_th_3sources}. From these quantities, all the marginal probabilities needed for the computation of the true values of the PID measures can be obtained and used for the evaluation of the estimation performances.

\begin{table}
\parbox{0.45\textwidth}{
\centering
\begin{tabular}{cccc}
\hline
\multicolumn{4}{c}{XOR2}                                                       \\ \hline
$X_1$ & \multicolumn{1}{c|}{$X_2$} & $p(Y=1,\textbf{X})$ & $p(Y=0,\textbf{X})$ \\ \hline
$<0$  & \multicolumn{1}{c|}{$<0$}  & $(b-1)(d-1)$        & 0                   \\
$\geq 0$  & \multicolumn{1}{c|}{$<0$}  & 0                   & $-b(d-1)$           \\
$<0$  & \multicolumn{1}{c|}{$\geq 0$}  & 0                   & $-(b-1)d$           \\
$\geq 0$  & \multicolumn{1}{c|}{$\geq 0$}  & $bd$                & 0                   \\ \hline
\multicolumn{4}{c}{}                                                           \\ \hline
\multicolumn{4}{c}{AND2}                                                       \\ \hline
$X_1$ & \multicolumn{1}{c|}{$X_2$} & $p(Y=1,\textbf{X})$ & $p(Y=0,\textbf{X})$ \\ \hline
$<0$  & \multicolumn{1}{c|}{$<0$}  & 0                   & $(b-1)(d-1)$        \\
$\geq 0$  & \multicolumn{1}{c|}{$<0$}  & 0                   & $-b(d-1)$           \\
$<0$  & \multicolumn{1}{c|}{$\geq 0$}  & 0                   & $-(b-1)d$           \\
$\geq 0$  & \multicolumn{1}{c|}{$\geq 0$}  & $bd$                & 0                   \\ \hline
\end{tabular}
\caption{Theoretical values of the joint probability of the target and source variables for the logic XOR2 and AND2 as function of the parameters $b$ and $d$.}
\label{prob_th_2sources}
}

\hfill

\parbox{0.45\textwidth}{
\centering
\begin{tabular}{ccccc}
\hline
\multicolumn{5}{c}{AND3}                                                               \\ \hline
$X_1$ & $X_2$ & \multicolumn{1}{c|}{$X_3$} & $p(Y=1,\textbf{X})$ & $p(Y=0,\textbf{X})$ \\ \hline
$<0$  & $<0$  & \multicolumn{1}{c|}{$<0$}  & 0                   & $-(b-1)(d-1)(f-1)$  \\
$<0$  & $<0$  & \multicolumn{1}{c|}{$\geq 0$}  & 0                   & $(b-1)(d-1)f$       \\
$<0$  & $\geq 0$  & \multicolumn{1}{c|}{$<0$}  & 0                   & $(b-1)d(f-1)$       \\
$<0$  & $\geq 0$  & \multicolumn{1}{c|}{$\geq 0$}  & 0                   & $b(d-1)(f-1)$       \\
$\geq 0$  & $<0$  & \multicolumn{1}{c|}{$<0$}  & 0                   & $-(b-1)df$          \\
$\geq 0$  & $<0$  & \multicolumn{1}{c|}{$\geq 0$}  & 0                   & $-b(d-1)f$          \\
$\geq 0$  & $\geq 0$  & \multicolumn{1}{c|}{$<0$}  & 0                   & $-bd(f-1)$          \\
$\geq 0$  & $\geq 0$  & \multicolumn{1}{c|}{$\geq 0$}  & $bdf$               & 0                   \\ \hline
\end{tabular}
\caption{Theoretical values of the joint probability of the target and source variables for the logic AND3 as function of the parameters $b$, $d$ and $f$.}
\label{prob_th_3sources}
}
\end{table}

\begin{figure*}
    \centering
    \includegraphics[scale=0.95]{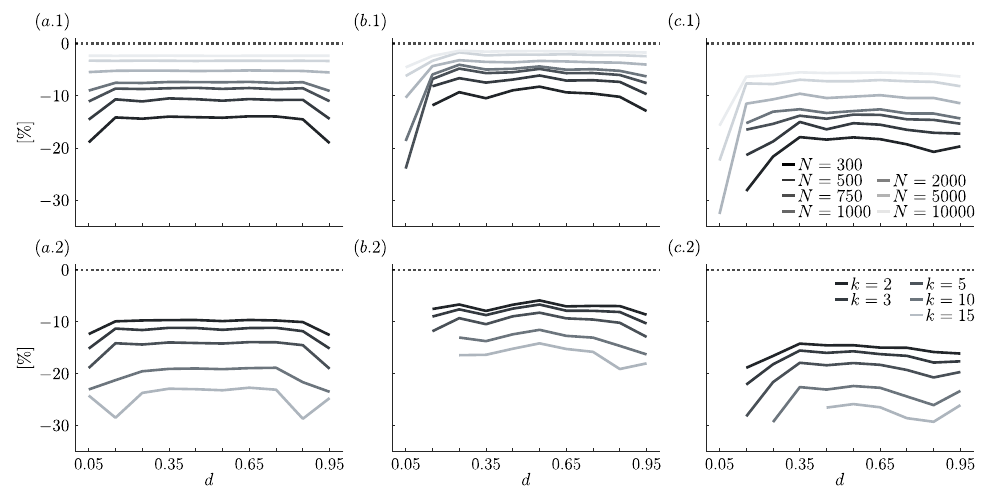}
    \caption{Average estimation of the percentage bias in computing the measure $I(Y;\textbf{X})$ on one-hundred realizations of the XOR-logic gate between two sources (panel \textit{a.}), of the AND-logic gate between two (panel \textit{b.}) and three (panel \textit{c.}) sources at varying the variable realizations length \textit{N} fixing $k=5$ (panel \textit{.1}) and the number of neighbors \textit{k} fixing $N=300$ (panel \textit{.2}).}
    \label{fig:bias}
\end{figure*}

It is known that the nearest-neighbor approach yields the greatest estimation bias when searching for samples in higher-dimensional space ~\cite{faes2014conditional}. In our framework of analysis, this is accounted for the overall mutual information between the target and all sources and for the atom at the top of the redundancy lattice. Accordingly, to report the worst-case estimation scenario, the computation of $I(Y;\textbf{X})$ has been chosen to assess the performance of our approach.
Figure~\ref{fig:bias} shows how the estimation bias (percent of the true value, average on one-hundred realization) is affected by the length $N$ of the realizations analyzed (panel \textit{.1}) and by the number of neighbors \textit{k} (panel \textit{.2}) used for the computation of $I(Y;\textbf{X})$ in the XOR2 (panel \textit{a.}), AND2 (panel \textit{b.}) and AND3 (panel \textit{c.}) gates. As expected from results previously reported on the performance of the nearest-neighbor approach~\cite{bara2024comparison}, the estimation bias overall decreases increasing the length $N$ of the data and using a lower number of searching neighbors $k$. Furthermore, as evidenced by the absence of data points in the plot, estimation of the PID terms is not always feasible when the target variable exhibits imbalanced values. This is particularly true when the data length is insufficient or when a higher number of neighbors is employed. Additionally, in such cases, the estimation is also affected by a higher percentage bias. Nevertheless, empirical corrections in the estimation procedure can be adopted to consider for these cases (e.g., forcing to zero the entropy conditioned to specific states of the target when such states are very unlikely to occur, so as to mimic the case of deterministic variables). Finally, a comparison of the results obtained for settings of different sizes (e.g., AND2 vs. AND3) reveals a considerable increase in the percentage bias with the addition of a third source.

\begin{figure*}
    \centering
    \includegraphics[scale=0.95]{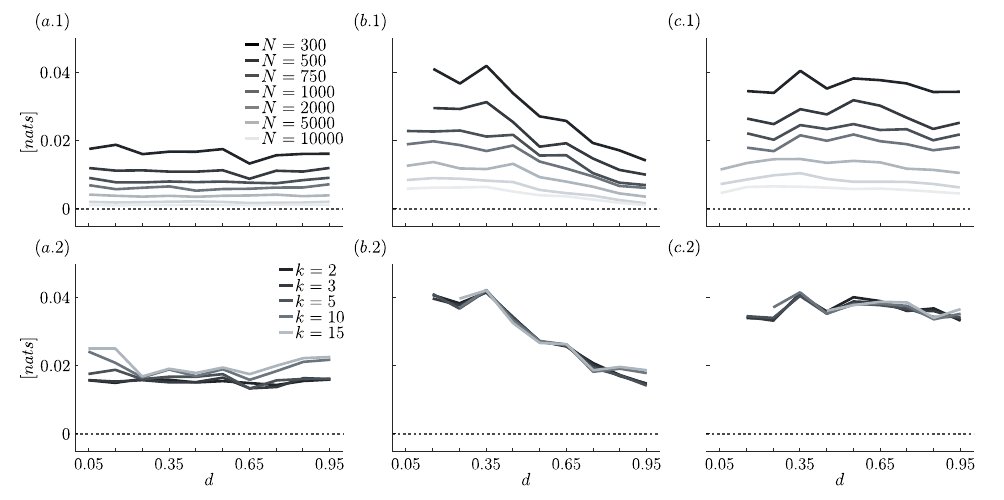}
    \caption{Standard deviation of the estimates of $I(Y;\textbf{X})$ on one-hundred realizations of the XOR-logic gate between two sources (panel \textit{a.}), of the AND-logic gate between two (panel \textit{b.}) and three (panel \textit{c.}) sources at varying the variable realizations length \textit{N} fixing $k=5$ (panel \textit{.1}) and the number of neighbors \textit{k} fixing $N=300$ (panel \textit{.2}).}
    \label{fig:std}
\end{figure*}

Figure~\ref{fig:std} depicts the standard deviation of the estimates of $I(Y;\textbf{X})$ evaluated on one hundred realizations of the XOR2 (panel \textit{a.}), AND2 (panel \textit{b.}) and AND3 (panel \textit{c.}) logic gates, and its dependence on the data length $N$ (panel \textit{.1}) and on the number of neighbors \textit{k} (panel \textit{.2}). The standard deviation appears strongly influenced by $N$ as it decreases as the length of the simulated realizations increases, while it remains overall unchanged when the number of searching neighbors $k$ is altered. Moreover, this indicator seems to be affected by the unequal assignment of samples to the target output.
In contrast with the estimation bias, it is not straightforward to identify in advance a measure that exhibits higher variation in the estimates, as this is not influenced by the dimensionality of the search space or the size of the investigated system (there are no evident differences in the comparison of AND2 and AND3 results). Therefore, we have decided to report the results for the measure $I(Y;\textbf{X})$ in this case too for consistency.

\section{Surrogate data analysis}
In this section, we develop a surrogate data analysis approach to assess the significance of the atoms of information resulting from applying the PID to the overall MI $I(Y;\textbf{X})$. The proposed approach is thus aimed at assessing the statistical significance of the partial information $I_\delta(Y;\alpha)$ for each of the atoms $\alpha$ of the redundancy lattice, but is then employed also to assess the significance of the coarse-grained PID terms (see, e.g., Fig. 1.b.2) by testing if at least one of the PI atoms combined to estimate such term contains significant information.

Following the estimation approach described in the main paper, the partial information measure for a generic collection of sources $\{X_s\}\in\alpha$ can be written as a combination of the specific information of quantities $I_\delta(Y=i;\alpha)$ related to a single value $i \in A_Y=\{1,\dots,Q\}$ of the target variable $Y$:
\begin{equation}
    \begin{split}
        &I_\delta(Y;\alpha)= I_\cap(Y;\alpha) - \sum_{\beta \prec \alpha} I_\delta(Y;\beta)=\\
        &=\sum_{i \in A_Y}p(Y=i)\left(I_\cap(Y=i;\alpha) - \sum_{\beta \prec \alpha} I_\delta(Y=i;\beta)\right)=\\
        &=\sum_{i \in A_Y}p(Y=i)I_\delta(Y=i;\alpha),
    \end{split}
    \label{sumatoms}
\end{equation}
where the redundant information of the atom $\alpha$ associated to a specific outcome $i$ of the target is obtained as (see also Eq. (4) of the main paper):
\begin{equation}
    I_\cap(Y=i;\alpha) = \min_{\{X_s\}\in \alpha}I(Y=i;X_s),
\end{equation}
and $\beta$ indicates the atoms preceding $\alpha$ in the redundancy lattice.
Eq. (\ref{sumatoms}) shows that $I_\delta(Y;\alpha)$ is obtained as the sum of $Q$ specific partial information quantities, whose significance can be tested by shuffling randomly the order of target samples. Specifically, the term $I_\delta(Y=i;\alpha)$ is computed on surrogate data and compared to the original measure, which is deemed as significant if its value is higher than the $\left(100-\frac{0.05}{Q \cdot m}\right)^{th}$ percentile of the distribution obtained on surrogate data, with $m$ the number of atoms of the redundancy lattice; thus, the partial information measure $I_\delta(Y;\alpha)$ is significant if at least one of the $Q$ summed terms $I_\delta(Y=i;\alpha)$ is significant. Thus, in the spirit of applying a Bonferroni correction, the division of the significance threshold by the number $Q$ of possible target outputs is needed to properly assess the significance of the PI $I_\delta(Y;\alpha)$. Moreover, an additional correction should be applied when looking at the statistical significance of any investigated coarse-grained term, further dividing the significance threshold by the number of atoms composing such term (a division to the number of atoms $m$ should be adopted to test the significance of the overall MI $I(Y;\textbf{X})$).
Indeed, for assessing the significance of $I(Y;\textbf{X})$ and of its decomposition terms (when $n>2$), it is essential to consider how these measure are obtained as combination of the partial information quantities of more lattice atoms.

\begin{figure*}
    \centering
    \includegraphics[scale=0.95]{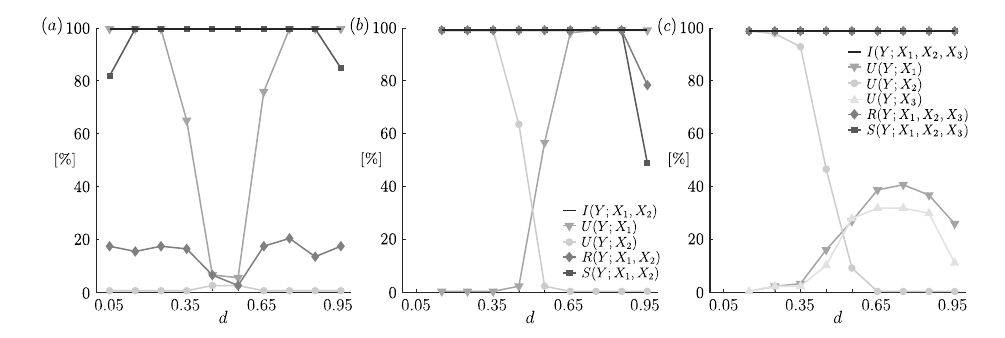}
    \caption{Percentage of realizations for which PID measures are statistically significant according to surrogate data analysis on simulated logic XOR (panel \textit{a}) and AND (panel \textit{b}) gates with 2 sources, and logic AND gate with 3 sources (panel \textit{c}).}
    \label{fig:surrogate}
\end{figure*}

The results of surrogate data analysis obtained on one hundred realizations of the logic Boolean gates simulated as described in the main paper are reported in Figure~\ref{fig:surrogate}; specifically, one thousand surrogate data have been generated for each of the realized gates shuffling randomly the observation of the binary target variable generated with a data length of $N=300$ (analysis performed with a number of searching neighbors $k=5$).
In all three simulated settings, the trends of significance of the PID terms overall follow those of the estimated measures, thereby assessing the feasibility of the introduced estimation approach. Nevertheless, in a system where the target variable is completely decoupled from the two source variables, as assumed in the generation of surrogate data, we can observe an overestimation of the terms assuming values close to zero (e.g., the redundancy term in the XOR gate or the unique contributions of $X_1$ and $X_3$ in the AND gates with three sources). Therefore, the bias obtained in the assessment of significance can be related to the variance of the estimates, which may also lead to a slight overestimation of these measures in some realizations.

\section{Physiological data details}

In this study, an historical database employed to investigate cardiovascular and cardiorespiratory dynamics in resting state and under postural stress has been analyzed~\cite{javorka2018towards}. It consists on synchronous acquisitions of electrocardiographic (ECG; CardioFax ECG-9620, NihonKohden, Japan), arterial blood pressure (ABP; Finometer Pro, FMS, Netherlands) and respiratory volume (RESP; RespiTrace, NIMS, USA) signals (sampling frequency of 1 kHz) on 61 young healthy subjects (37 females and 24 males, 17.5±2.4 years old). The cohort underwent an experimental protocol including two distinct conditions: a resting phase (REST) and a phase of postural stress (STRESS). During REST, subjects were instructed to lie in a supine position, while during STRESS they were passively tilted up at 45 degrees to evoke orthostatic stimulation. The Ethical Committee of the Jessenius Faculty of Medicine, Comenius University, Martin, Slovakia approved the acquisition protocol.

\begin{figure*}
    \centering
    \includegraphics[scale=0.95]{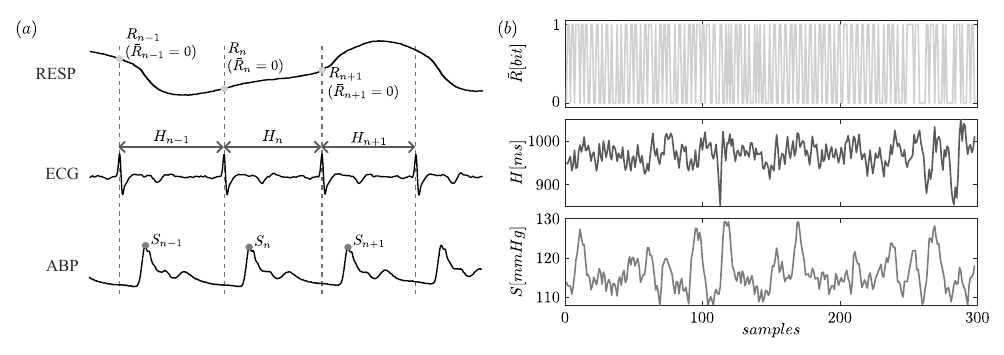}
    \caption{Schematic representation of the procedure for extracting the time series from the synchronously acquired respiratory (RESP), electrocardiographic (ECG) and arterial blood pressure (ABP) signals (panel \textit{a}), and resulting time series representative of respiratory phase ($\bar{R}$), heart period ($H$) and systolic arterial pressure ($S$) physiological dynamics, respectively (panel \textit{b}).}
    \label{fig:data}
\end{figure*}

From the acquired signals, stationary variability time series of heart period \textit{H}, systolic arterial pressure \textit{S} and respiration \textit{R} were extracted respectively (length of 300 samples) as the time distances between consecutive ECG R peaks, the local maxima of the ABP signal within each detected heart period, and the values of the respiratory signal sampled at each detected ECG R peak. 
To investigate the interplay between cardiovascular dynamics and the respiratory phase in response to postural stress by applying the method of analysis introduced in our work, it was necessary to conduct further processing of the data \cite{mijatovic2024assessing}. Specifically, the ventilatory phases of expiration and inspiration were assessed at each heartbeat discretizing the breathing time series and employed as target variable, i.e., $Y = \bar{R}_n$ with $\bar{R}_n = 0$ if $R_n > R_{n+1}$ and $\bar{R}_n=1$ if $R_n \le R_{n+1}$ respectively, while 2-lags cardiovascular dynamic patterns were used as representative of the source variables, i.e., $X_1=\left[H_{n} \: H_{n+1}\right]$ and $X_2=\left[S_{n} \: S_{n+1}\right]$. A graphical representation of the data processing is provided in Figure \ref{fig:data}.

\section{Toolbox description}

In this section, the Matlab toolbox \textit{model-free estimation of Partial Information Decomposition (mfPID)} is described. It is equipped with all the functions necessary to perform the analysis described in this study, with some demo scripts to demonstrate its capabilities. In addition to the functions for implementing the mixed discrete and continuous random variables estimation approach presented here, the toolbox also contains the functions for applying the original formulation working on discrete variables \cite{williams2010nonnegative}.

As shown in the list below, the functions of the mfPID toolbox can be grouped in five distinct categories: 
\paragraph*{\textbf{Functions for data manipulation}}
\begin{itemize}
    \item \textit{mfPID\_quantization}: uniform discretization of a series by using the binning approach (fixing the number of bins);
    \item \textit{mfPID\_nuquantization}: non-uniform discretization of a series (fixing the size of the quantization levels);
    \item \textit{mfPID\_B}: generation of the observation matrix indicating the number of samples in the past for each series;
    \item \textit{mfPID\_B\_lags}: generation of the observation matrix indicating the samples lag (both in the past and in the future);
    \item \textit{mfPID\_ObsMat\_V}, \textit{mfPID\_ObsMat\_lags}, \textit{mfPID\_SetLag}: form temporary observation matrices 
\end{itemize}
\paragraph*{\textbf{Functions for the computation of theoretical values of the PID measures}}
\begin{itemize}
    \item \textit{mfPID\_redundancy\_lattice\_2sources}: compute partial information quantities of the four atoms of the redundancy lattice from the specific information terms;
    \item \textit{mfPID\_2sources\_th}: compute theoretical PID measures for two discrete sources and one discrete target from joint probabilities;
    \item \textit{mfPID\_redundancy\_lattice\_3sources}: compute partial information quantities of the eighteen atoms of the redundancy lattice from the specific information terms;
    \item \textit{mfPID\_3sources\_th}: compute theoretical PID measures for three discrete sources and one discrete target from joint probabilities;
\end{itemize}
\paragraph*{\textbf{Function for the estimation of the PID measures}}
\begin{itemize}
    \item \textit{mfPID\_H}: estimate entropy for discrete multidimensional variable by using the frequentistic approach
    \item \textit{mfPID\_2sources\_dicrete}: estimate PID terms for two discrete sources and one discrete target from observation matrix;
    \item \textit{mfPID\_2sources\_mixed}: estimate PID terms for two continuous sources and one discrete target from the observation matrix;
    \item \textit{mfPID\_2sources\_mixed\_mex}: estimate PID terms for two continuous sources and one discrete target from the observation matrix (this makes use of closed mex functions);
    \item \textit{mfPID\_3sources\_dicrete}: estimate PID terms for three discrete sources and one discrete target from the observation matrix;
    \item \textit{mfPID\_3sources\_mixed}: Compute PID terms for three continuous sources and one discrete target from the observation matrix;
    \item \textit{mfPID\_3sources\_mixed\_mex}: estimate PID terms for three continuous sources and one discrete target from the observation matrix (this makes use of closed mex functions);
\end{itemize}
\paragraph*{\textbf{knn supporting functions} (mex files) \cite{lindner2011trentool}}
\begin{itemize}
    \item \textit{nn\_prepare}: generate a tree structure for nearest neighbor search;
    \item \textit{nn\_search}: use nearest neighbor search for a given query;
    \item \textit{range\_search}: find points within a range for a given query;
\end{itemize}
\paragraph*{\textbf{Surrogate data analysis functions}}
\begin{itemize}
    \item \textit{surrshuf}: generate surrogate data by random shuffling of the series samples.
\end{itemize}

The mfPID toolbox is freely available at the \href{https://sites.google.com/community.unipa.it/bitlab/software?authuser=0}{Biosignals and Information Theory Laboratory (BIT Lab) website} and on \href{https://github.com/ChiaraBara/mfPID_toolbox}{GitHub} along with the simulation setup of the Boolean logic AND-gate for two continuous sources and the real data application on one subject.

\bibliography{biblio.bib}

\end{document}